\documentclass[submission,copyright,creativecommons,noncommercial]{eptcs}
\providecommand{\event}{SR 2013} 
\usepackage{breakurl}             

\usepackage{amsmath,amssymb,latexsym,wasysym,dsfont,stmaryrd}
\usepackage{relsize}
\usepackage{xspace}
\usepackage{mymacros}
\usepackage{graphicx}
\usepackage{gastex}
\usepackage{tikz}
\usetikzlibrary{arrows,automata,shapes.geometric,decorations.pathmorphing,backgrounds,positioning,fit,petri}
\usepackage{wrapfig}
\usepackage{caption}
\usepackage{algorithm2e}

\title{The Complexity of Synthesizing Uniform Strategies}
\author{ 
Bastien Maubert
\institute{IRISA\\Universit\'e de Rennes 1\\Rennes, France}
\email{\quad bastien.maubert@irisa.fr}
\and
 Sophie Pinchinat
\institute{IRISA\\Universit\'e de Rennes 1\\Rennes, France}
\email{\quad sophie.pinchinat@irisa.fr}
\and
Laura Bozzelli
\institute{Facultad de Inform\'atica\\UPM\\ Madrid, Spain}
\email{laura.bozzelli@fi.upm.es}
}

\def\titlerunning{Synthesizing Uniform Strategies}
\def\authorrunning{B. Maubert, S. Pinchinat \& L. Bozzelli}
\begin{document}
\maketitle

\newif\ifdraft\drafttrue 

\ifdraft
\newcommand\bmaub[1]{\marginpar[{\color{cyan}\small\dbend}]{\color{cyan}\small\dbend}{\footnotesize \color{cyan}[#1 - \textbf{Bastien}]}}
\renewcommand\sp[1]{\marginpar[{\color{blue}\small\dbend}]{\color{blue}\small\dbend}{\footnotesize \color{blue}[#1 - \textbf{Sophie}]}}
\newcommand\bmaubchanged[1]{{\color{blue}{#1}}}
\newcommand\spchanged[1]{{\color{brown}{#1}}}
\else
\newcommand\bmaub[1]{}
\renewcommand\sp[1]{}
\newcommand\bmaubchanged[1]{#1}
\newcommand\spchanged[1]{#1}
\fi

\newtheorem{theorem}{Theorem}
\newtheorem{lemma}[theorem]{Lemma}
\newtheorem{corollary}[theorem]{Corollary}
\newtheorem{proposition}[theorem]{Proposition}
\newtheorem{definition}{Definition}
\newenvironment{proof}{\trivlist\item[\hskip \labelsep {\bf Proof}\enskip]}%
{\unskip\nobreak\hskip 2em plus 1fil\nobreak%
\fbox{\rule{0ex}{1ex}\hspace{1ex}\rule{0ex}{1ex}}%
\parfillskip=0pt \endtrivlist}

\begin{abstract}

  We investigate \emph{uniformity properties} of strategies. These
  properties involve sets of plays in order to express useful
  constraints on strategies that are not $\mu$-calculus
  definable. Typically, we can state that a strategy is
  observation-based.  We propose a formal language to specify
  uniformity properties, interpreted over two-player turn-based arenas
  equipped with a binary relation between plays. This way, we capture
  \eg games with winning conditions expressible in epistemic temporal
  logic, whose underlying equivalence relation between plays reflects
  the observational capabilities of agents (for example, synchronous
  perfect recall). Our framework naturally generalizes many other
  situations from the literature. We establish that the problem of
  synthesizing strategies under uniformity constraints based on
  regular binary relations between plays is non-elementary complete.
\end{abstract}

\section{Introduction}
In extensive infinite duration games, the arena is
represented as a graph whose vertices denote positions of players and
whose paths denote plays. In this context, a strategy of a player is a
mapping prescribing to this player which next position to select
provided she has to make a choice at this current point of the
play. As mathematical objects, strategies can be seen as infinite trees
 obtained by pruning the infinite unfolding of the
arena according to the selection prescribed by this strategy; outcomes
of a strategy are therefore the branches of the trees.

Strategies of players are not arbitrary in general, since players aim
at achieving some objectives. 
Infinite-duration game models have been
intensively studied for their applications in computer science
\cite{apt2011lectures} and logic \cite{booklncs2500}. First,
infinite-duration games provide a natural abstraction of computing
systems' non-terminating interaction \cite{alur2002alternating} (think of a communication protocol between a
printer and its users, or control systems). Second, infinite-duration
games naturally occur as a tool to handle logical systems for the
specification of non-terminating behaviors, such as for the
propositional $\mu$-calculus \cite{emerson91}, leading to a powerful
theory of automata, logics and infinite games \cite{booklncs2500} and
to the development of algorithms for the automatic verification
(``model-checking'') and synthesis of hardware and software systems.
In both cases, outcomes of strategies are submitted to
$\omega$-regular conditions representing some desirable property
of a system.

Additionally, the cross fertilization of multi-agent systems and
distributed systems theories has led to equip logical systems with
additional modalities, such as epistemic ones, to capture uncertainty
\cite{sato1977study,lehmann1984knowledge,fagin1991model,parikh1985distributed,ladner1986logic,halpern1989complexity}
, and more recently, these logical systems have been adapted to game
models in order to reason about knowledge, time and strategies
\cite{van2003cooperation,jamroga2004agents,dima2010model}. The whole
picture then becomes intricate, mainly because time and knowledge are
essentially orthogonal, yielding a complex theoretical universe to
reason about. In order to understand to which extent knowledge and
time are orthogonal, the angle of view where strategies are infinite
trees is helpful: Time is about the \emph{vertical} dimension of the
trees as it relates to the ordering of encountered positions along
plays (branches) and to the branching in the tree. On the contrary,
Knowledge is about the \emph{horizontal} dimension, as it relates
plays carrying, e.g., the same information.

As far as we know, this horizontal dimension, although extensively
studied when interpreted as knowledge or observation \cite{arnold02b,
  van2003cooperation,jamroga2004agents,chatterjee2006algorithms,alur2007model,dima2010model},
has not been addressed in its generality. In this paper, we aim at
providing a unified setting to handle it. We introduce the generic
notion of \emph{uniformity properties} and associated so-called
\emph{uniform strategies} (those satisfying uniformity properties). Some
notions of ``uniform'' strategies have already been used, e.g., in the
setting of strategic logics
\cite{van2001games,benthem2005epistemic,jamroga2004agents} and in the
evaluation game of Dependence Logic \cite{vaananen2007dependence},
which both fall into the general framework we present here.

We use a simple framework with two-player turn-based arenas and where
information lies in positions, but the approach can be extended to
other settings. Additionally, although uniformity properties can be
described in a set-theoretic framework, we propose the
logical formalism \RLTL which can be exploited to address fundamental
automated techniques such as the verification of uniformity properties
and the synthesis of uniform strategies -- arbitrary uniformity
properties are in general hopeless for automation. The formalism we
use combines the Linear-time Temporal Logic \LTL \cite{gabbay80} and
a new modality $\R$ (for ``for all related plays''), the semantics of
which is given by a binary relation between plays. Modality $\R$
generalizes the knowledge operator ``$K$'' of \cite{halpern1989complexity}
for the epistemic relations of agents in Interpreted Systems. The
semantic binary relations between plays are very little constrained:
they are not necessarily equivalences, to capture, \eg plausibility
(pre)orders one finds in doxastic logic \cite{hintikka1962knowledge},
neither are they knowledge-based, to capture particular strategies in
games where epistemic aspects are irrelevant. Formulas of the logic
are interpreted over outcomes of a strategy. The $\R$ modality allows
to universally quantify over all plays that are in relation with the
current play. Distinguishing between the universal quantification over
all plays in the game and the universal quantification over all the
outcomes in the strategy tree yields two kinds of uniform strategies:
the \emph{fully-uniform strategies} and the \emph{strictly-uniform
  strategies}.

As extensively demonstrated in \cite{maubertRapport2012}, uniform
properties turn out to be many in the literature: they occur in games
with imperfect information, in games with opacity conditions and more
generally with epistemic conditions, as non-interference properties of
computing systems, as diagnosability of discrete-event systems, in the
game semantics of Dependence Logic.

We investigate the automated synthesis of fully-uniform strategies,
for the case of finite arenas and binary relations between plays that
are rational in the sense of
\cite{berstel1979transductions}. Incidentally, all binary relations
that are involved in the relevant literature seem to follow this
restriction. In this context, two problems can be addressed: the
\emph{fully-uniform strategy problem} and the \emph{strictly-uniform
  strategy problem}, which essentially can be formulated as ``given a
finite arena, a finite state transducer describing a binary relation
between plays, and a formula expressing a uniformity property, does
there exist a fully-uniform (resp.\ strictly-uniform) strategy for
Player 1?''. From \cite{maubertRapport2012}, the fully-uniform
strategy problem is decidable but non-elementary -- since then we have
established that it is non-elementary hard. The algorithm involves an
iterated non-trivial powerset construction from the arena and the
finite state transducer which enables to eliminate innermost $\R$
modalities. Hence, the required number of iterations matches the
maximum number of nested $\R$ modalities of the formula expressing the
uniformity property. As expected, each powerset construction is
computed in exponential time. This procedure amounts to solving an
ultimate \LTL game, for which a strategy can be synthesized
\cite{pnueli89b} and traced back as a solution in the original problem.
The decidability of the strictly-uniform strategy problem is an open
question.

The rest of the paper is organized in five sections. In
Section~\ref{sec-preliminaries}, we present the standard material
two-player turn-based arenas. We set up the framework and define
uniform strategies in Section~\ref{sec-uniformproperties}, and we
illustrate the notion with two examples in Section~\ref{sec-literature}.
Finally
in Section~\ref{sec-complexity}, we give tight complexity bounds for the fully-uniform
strategy problem, and we discuss future work in Section~\ref{sec-discussion}.

\section{Preliminaries} 
\label{sec-preliminaries}

We consider two-player turn-based games that are played on graphs with
vertices labelled with propositions.
These propositions represent the relevant information for the uniformity properties one
wants to state.
From now on and for the rest of the paper, we let $\AP$ be an
infinite set of \emph{atomic propositions}.

An \emph{arena} is a structure $\ga=(V,E,v_0,\val)$ where $V=V_1\uplus
V_2$ is the set of \emph{positions}, partitioned between
positions of Player 1 ($V_1$) and those of Player 2 ($V_2$),
$E\subseteq (V_1\times V_2) \cup (V_2\times V_1)$ is the set of
\emph{edges}, $v_0\in V$ is the \emph{initial position} and
$\val:V\rightarrow
 \parti{\AP}$ is a \emph{valuation function}, mapping each position to the finite
 set of atomic propositions that hold in this position. 
%
$\PlaysFin$ and $\PlaysInf$ are, respectively, the set of finite and
infinite \emph{plays}.
For an infinite play $\pi=v_0v_1\ldots$ and $i\in
 \mathbb N$, $\pi[i]:=v_i$ and $\pi[0,i]:=v_0\ldots v_i$. 
For a finite play $\rho=v_0v_1\ldots v_n$, $\last(\rho)=v_n$. 
%

A \emph{strategy} for Player $1$  is a partial function $\sigma
: \PlaysFin \rightarrow V$   that maps a finite play ending in $V_1$
to the next position to play. Let $\sigma$ be a strategy for
Player $1$. We say that  a play $\pi\in\PlaysInf$ 
is \emph{induced by} $\sigma$ if for all $i\geq 0$ such that
$\pi[i]\in V_1$, $\pi[i+1]=\sigma(\pi[0,i])$, and the \emph{outcome
  of} $\sigma$, noted $\out(\sigma)\subseteq \Plays_\omega$, is the
set of all infinite plays that are induced by $\sigma$. Definitions
are similar for Player~$2$'s strategies.

\section{Uniform strategies} 
\label{sec-uniformproperties}

We define the formal language \RLTL to specify
uniformity properties. 
This language enables to express properties of the dynamics of plays,
and resembles the Linear Temporal Logic (\LTL)
\cite{gabbay80}. However, while \LTL formulas are evaluated on
individual plays (paths), we want here to express properties on
``bundles'' of plays. To this aim, we equip arenas with a
binary relation between finite plays, and we enrich the logic with a
modality $\R$ that quantifies over related plays, the intended
meaning of ``$\R\phi$ holds in $\rho$'' being ``$\phi$ holds
in every play related to $\rho$''.

\label{sec-syntax}
The syntax of \RLTL is similar to that of linear temporal logic with
knowledge \cite{halpern1989complexity}.
However, we use $\R$ instead of the usual knowledge operator $K$ to
emphasize that 
it need not be interpreted in terms of knowledge in general, but merely
as a way to state properties of bundles of plays. The syntax is:

\[\varphi,\psi ::= p\mid \neg \varphi \mid \varphi\wedge \psi  \mid \fullmoon \varphi \mid \varphi\until\psi
  \mid \R \varphi \mbox{\hspace{1cm}}  p\in\AP\]


\label{sec-semantics}
 Consider an arena $\ga=(V,E,v_0,\val)$ and a rational relation $\leadsto\;\subseteq
\PlaysFin\times\PlaysFin$. A formula $\phi$ of \RLTL is evaluated at
some point $i\in\mathbb N$ of an infinite play $\pi\in\PlaysInf$,
within a \emph{universe} $\Pi\subseteq \PlaysInf$.
The semantics is given by induction over formulas. 

\vspace{.3cm}
{\centering $\begin{array}{ll}
  \Pi,\pi,i\models p \; \mbox{ if }\; p\in \val(\pi[i]) &
  \Pi,\pi,i\models \neg \varphi \; \mbox{ if }\; \Pi,\pi,i\not\models \varphi \\
\Pi,\pi,i\models \varphi \wedge \psi \; \mbox{ if }\;  \Pi,\pi,i \models
\varphi \mbox{ and } \Pi,\pi,i\models \psi \mbox{~~~~~~} &
  \Pi,\pi,i\models \fullmoon \varphi \; \mbox{ if } \; \Pi,\pi,i+1 \models \varphi \\
\multicolumn{2}{l}{\Pi,\pi,i\models \varphi \until \psi \; \mbox{ if }
  \;\mbox{there is }j\geq i \mbox{ such that }\Pi,\pi,j\models \psi \mbox{ and for all }i\leq k < j,\; \Pi,\pi,k \models \varphi}\\
\multicolumn{2}{l}{  \Pi,\pi,i\models \R \varphi \; \mbox{ if } \; \mbox{for all }\pi'\in\Pi,j\in \mathbb N \mbox{ such that }\pi[0,i]\leadsto\pi'[0,j]
  ,\;\Pi,\pi',j\models \varphi}
\end{array}$

}
\vspace{.3cm}

\noindent From this semantics, we derive two notions of uniform strategies,
which differ only in the universe the $\R$ modality quantifies over:
$\out(\sigma)$ or $\PlaysInf$ (with the latter, related plays not
induced by the strategy also count). The motivation for these two
definitions is clear from \cite{maubertRapport2012} where many examples
from the literature are given.

\begin{definition}
\label{def-unifstrat}
  Let $\ga$ be an arena, $\leadsto$ be a rational relation and $\varphi$
  be an \RLTL formula.   A strategy $\sigma$ is:
 
{\centering $(\leadsto,\varphi)$-\emph{strictly-uniform} if
for all $\pi\in \out(\sigma)$, $\out(\sigma),\pi,0\models \varphi$,

$(\leadsto,\varphi)$-\emph{fully-uniform} if
for all $\pi\in \out(\sigma)$, $\PlaysInf,\pi,0\models \varphi$.

}

\end{definition}

\section{Concrete examples}
\label{sec-literature}
In this section we illustrate our notions of strictly and fully uniform
strategies defined in the previous section with the examples of
observation-based strategies in games with imperfect information, and
games with opacity condition.  



\subsection{Observation-based strategies} 
\label{sec-observation}

Games with imperfect information, in general, are games
in which some of the players do not know exactly what is the current
position of the game. Poker is an example of imperfect-information
game: one does not know which cards her opponents have in hands.
One important aspect of imperfect-information games is that not every
strategy is ``playable''.
Indeed, a player cannot plan to play differently in  situations that she
is unable to distinguish.  This is why players are required to use
strategies that select moves uniformly over
observationally equivalent situations. This kind
of strategies is sometimes called \emph{uniform strategies} 
 in the
community of strategic logics (\cite{van2001games,benthem2005epistemic,jamroga2004agents}), or \emph{observation-based strategies}
 in the community of computer-science oriented
game theory (\cite{chatterjee2006algorithms}). In fact, all the
additional complexity of solving imperfect-information games, compared to
perfect-information ones, lies in this constraint put on strategies. 
We  show that the notion of
observation-based strategy, and hence the essence of games with
imperfect information, can be easily embedded in our notion of
uniform strategy. 
In two-player imperfect-information games as studied
for example in
\cite{reif84,chatterjee2006algorithms,berwanger2008power}, Player
$1$ only partially observes the positions of the game, such that
some positions are indistinguishable to her, while Player $2$ has
perfect information (the asymmetry is due to the focus being on the
existence of strategies for Player $1$). Arenas are labelled
directed graphs together with a finite set of \emph{actions} $\Act$, and in each round, if
the position is a node $v$, Player $1$ chooses an available action $a$, and
Player $2$ chooses a next position $v'$ reachable from $v$ through
an $a$-labelled edge.

We equivalently define this framework in a manner that fits our
setting by putting Player $1$'s actions inside the positions. We
have two kinds of positions, of the form $v$ and of the form
$(v,a)$. In a position $v$, when she chooses an action $a$, Player
$1$ actually moves to position $(v,a)$, then Player $2$ moves
from $(v,a)$ to some $v'$.  So an imperfect-information game arena is
a structure $\mathcal G_{imp}=(\mathcal G,\sim)$ where $\mathcal
G=(V,E,v_0,\val)$ is a two-player game arena with positions
in $V_1$ of the form $v$ and positions in $V_2$ of the form $(v,a)$.
We require that $vE (v',a)$ implies
$v=v'$, and $v_0\in V_1$.
For a position $(v,a)\in V_2$,  we note $(v,a).act :=a$.
 We assume that $p_1\in\AP$, and for every
 action $a$ in $\Act$, $p_a\in\AP$. $p_1$ holds in positions belonging
to Player~1, and $p_a$ holds in positions of Player~2 where the last
action chosen by Player~1 is $a$: $\val(v)=\{p_1\}$ for $v\in V_1$, $\val(v,a)=\{p_a\}$ for
$(v,a)\in V_2$. Finally, $\sim\;\subseteq V_1^2$ is an observational
equivalence relation on positions, that relates positions
indistinguishable for Player~$1$. We define its extension $\simeq$ to finite
plays: $v_0 (v_0,a_1) v_1 \ldots  (v_{n-1},a_n) v_n \simeq v_0 (v_0,a'_1)
v'_1 \ldots  (v'_{n-1},a'_n) v'_n$ if for all $i>0$, $v_i\sim v'_i$
and $a_i=a'_i$.


We add the classic
requirement that the same actions must be available in indistinguishable positions: for all
$v,v'\in V_1$, if $v\sim v'$ then $vE (v,a)$ if, and only if,
$v'E (v',a)$. In other words, if Player~$1$ has different options, she
can distinguish the positions.


\begin{definition} 
A strategy $\sigma$ for Player $1$ is \emph{observation-based} if 
for all $\rho, \rho' \in v_0(V_2V_1)^*$, $\rho\approx\rho'$  implies $\sigma(\rho).act=\sigma(\rho').act$.
\end{definition}

 We define the formula
\[\text{\tt
  SameAct}:=\G(p_1\rightarrow\bigvee\limits_{a\in\Act}\!\!\!\R \fullmoon
p_a)\]
which, informally, expresses that whenever it is Player $1$'s turn to play, there is an action $a$ that is played
in every equivalent finite play.

\begin{proposition}
A strategy $\sigma$ for Player $1$ is observation-based iff it is $(\approx,\text{\tt SameAct})$-strictly-uniform.
\end{proposition} 

Here we have to make use of the notion of strict uniformity, and not the
full uniformity. Indeed, after a finite play $\pi[0,i]$ ending in $V_1$,
we want to enforce that in all
equivalent prefixes of infinite plays \emph{that conform to the strategy
  considered}, Player~$1$ plays the same action. It would obviously
make no sense to enforce the same on equivalent prefixes of  every
possible play in the game, which encompass all possible behaviours of Player~$1$.



Notice that in order to embed the case of players with different memory
abilities, \eg imperfect-recall, one
would just have to replace $\approx$ with the appropriate relation.

For the moment we have not mentioned any winning condition. For a strategy,
being
$(\approx,\text{\tt SameAct})$-strictly-uniform only characterizes that
it is ``playable'' for a player with imperfect
information, but it does not characterize the outcome of this
strategy. 
However, if one considers a game with imperfect information in which
the winning condition for Player~$1$ is an LTL
formula $\phi$, then the set of $(\approx,\text{\tt SameAct}\wedge
\phi)$-strictly-uniform strategy is exactly the set of
winning observation-based strategy.
 
When talking about knowledge and strategic abilities,
the question of \emph{objective} vs \emph{subjective} ability should
be raised (see \cite{jamroga2011comparing}). The difference is basically whether a
strategy is defined only on ``concrete'' plays,
starting from the initial position, or if it has to be defined on all ``plays'' starting from
any position the player 
confuses with the initial one. In the setting presented here, 
 the initial position is part of the description of the arena,
hence players are assumed to know it and all plays considered start from
this position. But in order to model in this setting the case of Player 1
not knowing the initial position, one could
add a fresh artificial initial position $v_0'$, from which no matter
the action Player 1 chooses, Player 2
can move to any position that Player 1 confuses with $v_0$. Then, for
a winning condition $\phi\in\LTL$,
the existence of an observation-based winning strategy for Player 1 from $v_0$
(resp. $v_0'$) would denote
objective (resp. subjective) ability to enforce $\phi$.



\subsection{Games with opacity condition}
\label{sec-opacity}


Games with opacity condition,  studied in \cite{maubert2011opacity}, are based on two-player
imperfect-information arenas, with Alice having perfect information
as opposed to Bob who partially observes positions. In such games, some positions
are ``secret'' as they reveal a critical information that Bob aims at
discovering. We are interested in Alice's ability to prevent Bob from
``knowing'' the secret, in the epistemic sense.

More formally, assume that a proposition $p_S\in \AP$ represents the secret. Let $\mathcal G_{inf}=(\mathcal G,\sim)$ be an
imperfect-information arena as described in Section~\ref{sec-observation},
with a distinguished set of positions $S\subseteq V_1$ that denotes
the secret. Bob is Player~$1$ as he has imperfect information,
and Alice is Player~$2$. 
Letting $\mathcal G=(V,E,v_I,\val)$, we require that
$\val^{-1}(\{p_S\})=S$ (positions labeled by $p_S$ are exactly positions $v \in S$).  
For a finite play $\rho$ with $last(\rho) \in
  V_1$, Bob's \emph{information set} or \emph{knowledge} after $\rho$ is $I(\rho):=\{last(\rho')\mid \rho'\in\PlaysFin,
  \rho\approx\rho'\}$.
It is the set of all the positions he considers
possible after observing $\rho$.
An infinite play is winning for Bob if there exists a finite prefix $\rho$ of this play
whose information set is contained in $S$, \ie $I(\rho)\subseteq S$, otherwise Alice wins.
It can easily be shown that:
\begin{proposition}
\label{theo-opacity}
A strategy $\sigma$ for Alice is winning if, and only if, $\sigma$ is $(\approx,\G \neg \R p_S)$-fully-uniform.
\end{proposition}

Here we are interested in Alice's strategies and Bob's
knowledge. Since Bob only partially observes what Alice is playing,
some plays that are not brought about by Alice's strategy are considered
possible by Bob. Full
uniformity is therefore the right notion to capture correctly Bob's knowledge.  

Here again, to model different memory and observational abilities of Bob, one
can use the appropriate binary relation, provided it is
rational. Also, notice that though we chose to illustrate our framework with opacity
aspects, any winning condition that is expressible by a formula
of the epistemic linear temporal logic with one knowledge operator would fit in our setting.

\section{Synthesizing fully-uniform strategies}
\label{sec-complexity}
In this section, we investigate the complexity of synthesizing a
fully-uniform strategy. We first consider the associated decision
problem, called here the \emph{fully-uniform strategy problem}: given
a uniform property $\phi \in \RLTL$, a finite arena
$\ga=(V,E,v_0,\val)$, and a finite state transducer $T$ over alphabet
$V$ representing a rational binary relation between plays (see \cite{berstel1979transductions}), does there exist a
$([T],\phi)$-fully-uniform strategy in $\ga$, where $[T]$ is the
binary relation denoted by $T$.

\begin{definition}
\label{def-depth}
For a formula $\phi\in\RLTL$, the \emph{$\R$-depth} of
$\phi$, written $\depth(\phi)$, is the maximum number of nested $\R$
modalities in $\phi$. For each $k\in\setn$, we let $\RLTLn{k}\egdef\{\phi\in\RLTL
\mid \depth(\phi)=k\}$.
\end{definition}

 
\begin{theorem}
\label{prop1}
The fully-uniform strategy problem for formulas ranging over $\bigcup_{k\leq n}\RLTLn{k}$
is $n$-\EXPTIME-complete for $n> 2$, and \2EXPTIME-complete for
$n\leq2$.
\end{theorem}

The proof for the upper bounds in Theorem~\ref{prop1} can be found
in \cite{maubertRapport2012}, in which we devise a decision procedure
based on a powerset construction which simulates the execution of the
transducer along plays in the arena, enabling the computation of
information sets. Dealing with information sets enables us to perform
$\R$-modalities elimination, yielding a reduction of the initial
problem to solving some \LTL game. The procedure is however non-elementary
since it requires one powerset construction per nesting of
$\R$-modalities. The proof for the matching lower bounds is a direct reduction from the
word problem for $\exp[n]$-space bounded alternating Turing Machines,
which is $(n+1)$-\EXPTIME{} complete. Due to lack of space, it is omitted here. 

\begin{corollary}
\label{coro1}
The fully-uniform strategy problem is non-elementary complete.
\end{corollary}

Regarding the synthesis problem, the procedure of \cite{pnueli89b} for
solving the terminal \LTL game in the decision procedure of
Theorem~\ref{prop1} is an effective construction of a winning strategy
when it exists. This strategy provides a fully-uniform strategy of
the initial game, by means of a transducer mapping plays of the
initial game to plays in the terminal game. This transducer itself is
straightforwardly built from the arena of the last game itself. 

%

\section{Discussion}
\label{sec-discussion}

We are currently working on sufficient conditions on the binary
relation between plays to render the fully-uniform strategy synthesis problem elementary. It
appears that being an equivalence relation is not
enough, but if moreover the relation verifies a weak form of \emph{no
  learning} property (see \cite{halpern2004complete}), the problem
seems to be elementary. 
Concerning the strictly-uniform strategy problem, we conjecture
undecidability in general, but we are investigating interesting subclasses of rational
relations that make the problem decidable. 

It would then be interesting to extend the language to the case of $n$
modalities $\R_i$ with $n$ relations $\leadsto_i$. Also,
 the difference between the fully-uniform semantics and
the strictly-uniform one could be at the level of modalities instead
of the decision problems level. In Section~\ref{sec-observation} we
have seen that uniformity properties can represent \emph{uniformity constraints}
on the set of elegible strategies, and in Section~\ref{sec-opacity} we
have seen how they can represent \emph{epistemic winning
  conditions}. However, while some properties require strict
uniformity, others require full uniformity. Allowing to use both kinds of
modalities in a formula would enable, for example, to express that a strategy must
both be winning for some condition on the opponent's knowledge (with a
fully-uniform modality, see Section~\ref{sec-opacity}), and to be
observation based for the player considered (with a strictly-uniform
modality). A formula of the following kind could be used for a variant of games
with opacity condition where Alice would also have imperfect
information (note that the arenas should be modified, and we assume
that $p_2$ would mark positions where Alice has to choose an action):

\[\phi := \underbrace{\G(p_2\rightarrow\bigvee\limits_{a\in\Act}\!\!\!\R_{Alice}^{strictly}
  \fullmoon p_a)}_{\mbox{Observation-based constraint}} \; \wedge \;  \underbrace{\G \neg \R_{Bob}^{fully} p_S\rule[-13.5pt]{0pt}{25pt}}_{\mbox{Winning
  condition}}\]

In a next step, we would like to consider how our framework 
adapts if we take as base language the one of Alternating-time
Temporal Logic \cite{alur2002alternating} instead of LTL, so as to
obtain an Alternating-time Temporal Epistemic Logic-like
language. It would enable us to express the existence of uniform
strategies directly in the logic, and not only at the level of
decision problems as it is the case for now. This step will require to
pass from the two-player turn-based arenas considered so far to multiplayer concurrent game
structures, that are ATL models, but the definitions should adapt
without difficulties. However we should be cautious in generalizing
these notions as undecidability will easily be attained.  


\bibliographystyle{eptcs}

\end{document}